# Real-time observation of dynamics in rotational molecular wave packets by use of "air laser" spectroscopy

Bin Zeng,[1] Wei Chu,[1] Guihua Li,[1,2] Jinping Yao,[1] Haisu Zhang,[1,2] Jielei Ni,[1] Chenrui Jing,[1,2] Hongqiang Xie,[1,2] and Ya Cheng[1,*]

[1]State Key Laboratory of High Field Laser Physics, Shanghai Institute of Optics and Fine Mechanics, Chinese Academy of Sciences, P.O. Box 800-211, Shanghai 201800, China

[2]University of Chinese Academy of Sciences, Beijing 100049, China

[*]Email: ya.cheng@siom.ac.cn

**Abstract**

Molecular rotational spectroscopy based on strong-field-ionization-induced nitrogen laser is employed to investigate the time evolution of the rotational wave packet composed by a coherent superposition of quantum rotational states created in a field-free molecular alignment. We show that this technique uniquely allows real-time observation of the ultrafast dynamics of the molecular rotational wave packet. Our analysis also shows that there exist two channels of generation of the nitrogen laser, shedding new light on the population inversion mechanism behind the air laser generated by intense femtosecond laser pulses.

# 1. Introduction

Field-free alignment of linear molecules [1, 2] has found widespread applications including strong-field ionization [3, 4], high-order harmonic generation [5-7], molecular imaging [8, 9], and so on. During the interaction between ultrafast laser and molecules, a rotational wave packet is created through the non-resonant impulsive Raman excitation. After the extinction of the laser pulse, the field-free evolution of the rotational wave packet results in periodic revivals of molecular alignment and anti-alignment. The degree of the alignment is usually evaluated by a time-dependent parameter $\langle cos^2\theta \rangle(\tau)$ ($\theta$ is the angle between the molecular axis and the field polarization). The time evolution of the rotational wave packet can be measured via Coulomb explosion imaging [10], Kerr effect birefringence induced techniques [11, 12], high-order harmonic generation [13, 14], and so on. It should be stressed that the rotational wave packet is a broad, coherent super-position of discrete quantum rotational states. Each quantum rotational state is characterized with a specific angular momentum that is determined by the rotational quantum number. Previous investigations often focused on the behaviors of the overall rotational wave packets, i.e., the super-positions of many quantum rotational states of different angular momenta, whereas observation of the evolution of the rotational state distribution has been proved difficult.

Recently, Yao et al. [15, 16] reported the generation of nitrogen laser by strong-field-ionization of nitrogen molecules with femtosecond laser pulses. In a typical pump-probe scheme, the nitrogen laser can be generated with an 800 nm pump pulse, followed by a 400 nm probe pulse which is produced by frequency-doubling of the output pulse [16]. The 800 nm pump beam first ionizes the nitrogen molecules, causing a population inversion between the excited state $B^2\Sigma_u^+$ and the ground state $X^2\Sigma_g^+$ of the nitrogen ions $N_2^+$. Then the probe beam acts as a seed which will be amplified to generate the narrow-bandwidth, coherent nitrogen lasing emission. It is noteworthy that although the existence of population inversion has been confirmed with systematic experimental investigations, the mechanism behind the establishment of the inverted population within an ultra-short period of only ~100 fs is still under hot debate. In fact, several models have been proposed to understand the physics of such air laser, including stimulated emission based on seed amplification in a population inverted system, four-wave mixing, and seed-triggered superradiance [17-20]. Nevertheless, none of these models has been completely confirmed and widely accepted.

Interestingly, in the nitrogen lasing emission experiment, the pump laser pulses not only ionize the nitrogen molecules, but also induce an efficient alignment of both the molecular ions and the neutral molecules because of the high peak intensity of the pump pulses. The molecular rotational wave packet is a coherent superposition of rotational eigenstates excited from the initial state,

$$|\Psi(t)\rangle = \sum_J a_{J,M}(t)e^{-iE_Jt/\hbar}|J,M\rangle, \qquad (1)$$

where $E_J = hcBJ(J + 1)$ are the rotational eigenenergies and $|J, M\rangle$ the rotational states. $B$ is the rotational constant and $c$ is the speed of light. Remarkably, Zhang et al. recently found that the rotational coherence can faithfully encode its characteristics into the "air laser" spectrum. In such a case, the 391 nm lasing signal can be expressed by $e^{-\tau/\Gamma} \times \langle cos^2\theta \rangle(\tau)$, where the exponential term $e^{-\tau/\Gamma}$ accounts for the decay of the population inversion between the $B^2\Sigma_u^+$ and $X^2\Sigma_g^+$ states, and $\langle cos^2\theta \rangle(\tau)$ describes how the nitrogen lasing emission depends on the molecular rotational wavepacket [21]. However, in this work, only the integrated signal of all the lasing lines in the *P*-branch or *R*-branch was investigated with the pump-probe measurement, which cannot provide the dynamic information on evolution of the rotational state distribution in the wave packet of molecular ions of nitrogen.

In this paper, for the first time to our knowledge, we report on the real-time observation of the temporal evolution of the rotational state distribution with strong-field-ionization induced nitrogen lasing technique. The nitrogen lasing provides an ideal tool in this experiment. First of all, the pulse durations of the pump and the probe lasers are both of ~100 fs level. Pump-probe measurements based on these pulses can provide a sufficient temporal resolution to reveal the rotational dynamics of many kinds of molecules such as the nitrogen molecule investigated in this work. Second, the bandwidth of the strong-field-ionization induced molecular lasing spectrum is typically on the scale of ~0.1 nm, providing a sufficient spectral resolution to distinguish the different rotational states [15, 21]. In our point of view, the combination of these advantages opens new possibilities for investigation of strong field molecular physics.

## 2. Experiment

The experimental arrangement is similar to that used in our recent work reported in Refs. [20, 21]. The femtosecond laser beam (1 kHz, 800 nm, ~40 fs) produced by a commercial Ti:sapphire laser system (Legend Elite-Duo, Coherent, Inc) was divided into two arms. One with a pulse energy of 2.57 mJ served as the pump to ionize and align the nitrogen molecules, and the other, after being frequency doubled by a 0.2 mm-thick BBO crystal, served as the probe to generate the nitrogen lasing. The polarization of the probe pulse was set to be parallel to the pump pulse. The time delay between the pump and the probe pulse was controlled by a motorized linear translation stage with a temporal resolution of 16.7 fs. After combined by a dichroic mirror, the pump and the probe beams were focused by a fused-silica lens with a focal length of 40 cm into a vacuum chamber filled with nitrogen gas at a pressure of ~4 mbar. The lasing emissions generated in the chamber were then collimated by a f=30 cm lens and recorded by a 1200-grooves/mm grating spectrometer (Andor Shamrock 303i).

## 3. Fast oscillations in pump-probe measurements of laser signals from individual rotational levels

Figure 1 shows a typical spectrum of the nitrogen lasing emission which corresponds to the 0-0 band of the first negative system ($B^2\Sigma_u^+ \rightarrow X^2\Sigma_g^+$) of $N_2^+$ [22]. One can clearly see that both the rotational P-branch ($|J\rangle \to |J+1\rangle$) band and the R-branch ($|J\rangle \to |J-1\rangle$) lines appear in Fig. 1. The intensity of P-branch as a function of the time delay between the pump and the probe is plotted in Fig. 2(a). The zero time is defined as the pump and the probe pulses overlap. In principle, both the P- and R-branches are composed of a series of lines. However, in the P-branch, these lines are all very closely spaced, thus they merge into one peak due to the limited resolution of our spectrometer. For this reason, the lasing emission in the P-branch cannot provide useful information to distinguish the contributions of individual rotational quantum states with different rotational quantum number. As indicated in Fig. 2(b), only the distribution of the rotational states can be obtained by performing the Fourier-transform of the curve in Fig. 2(a). However, in the R-branch, the spectral lines from different rotational quantum number $J$ are well resolved because these lines are distributed more sparsely in the spectrum as compared to the P-branch lines. The R-branch band can thus be employed as a probe to investigate the real-time dynamics of the rotational state distribution. The intensities of R-branch lines from $J = 5$ to $J = 17$ states as functions of the time delay are plotted in Fig. 3. The insets in the figure show the details for a window of the time delay between 5 ps to 10 ps. Similar to the behavior of the overall signal of P-branch band, the signal of each line in the R-branch also decays exponentially with the increase of the time delay between the pump and the probe pulses, mainly because of the limited lifetime of the population inversion in the excited system. However, unlike the P-branch curve in Fig. 2(a), the curves in Fig. 3 show regular rapid oscillations with periods of sub-picosecond and regular slow oscillations with periods of several picoseconds.

It can be seen that the period of the rapid oscillation becomes shorter with increase of the $J$ value. By performing the Fourier-transform of these curves, as shown in Fig. 4, we found that the frequencies of the oscillations can be expressed by $(4J+6)B_Bc$, which is due to the quantum beat between the two adjacent rotational states $|J\rangle$ and $|J+2\rangle$,

$$\Omega_1 = \frac{E_{J+2,B} - E_{J,B}}{h} = (4J+6)B_Bc, \qquad (2)$$

where $B_B$ = 2.073 cm$^{-1}$ [23] is the rotational constant of the electronic state $B^2\Sigma_u^+$. These fast oscillations can be understood as a result of the temporal modulation of the population in the rotational *J* state caused by a coupling between the *J* and *J+2* states. The coupling of the two rotational states occurs through a resonant Raman-like process enabled by the 400 nm probe pulse. Since the 400 nm probe pulse has a broad spectrum which covers the *R*-branch of 391 nm fluorescence lines, the molecular ion in the excited state $B^2\Sigma_u^+ |J+2\rangle$ state can decay to $X^2\Sigma_g^+ |J+1\rangle$ by stimulated emission induced by the 400 nm probe pulse, and then be excited to $B^2\Sigma_u^+ |J\rangle$ by absorbing a photon from the probe pulse. During this coherent process, the phase of $B^2\Sigma_u^+ |J+2\rangle$ state can be transferred to $B^2\Sigma_u^+ |J\rangle$. The interference between $B^2\Sigma_u^+ |J\rangle$ state originally formed in the intense pump laser field and that formed by the above-mentioned resonant Raman-like process leads to the quantum beat in the population in $B^2\Sigma_u^+ |J\rangle$, which in turn results in the beat observed in the lasing line corresponding to $B^2\Sigma_u^+ |J\rangle$ (e.g., see, Figs. 3 and 4). This is analogous to the two-pulse alignment process in which a varying time delay between the two pulses can lead to modulation of the populations of different $J$ states [24].

## 4. Slow oscillations in pump-probe measurements of laser signals from individual rotational levels

Now let us focus on the slow oscillations in Fig. 3. We can see that the period of the slow oscillation first increases with the rotational quantum number $J$ until $J$ reaches ~11. Then the period of the slow oscillation becomes shorter as the rotational quantum number increases further. The frequencies of the slow oscillations as functions of the rotational quantum number are plotted by the stars in Fig. 5. We can see that the frequency of the slow oscillation decreases proportionally to the rotational quantum number $J$ when $J < 11$, whereas increases proportionally to the rotational quantum number $J$ when $J > 11$.

The slow oscillation can be explained by considering a cascade process involving first absorption of a *P*-branch photon followed by emission of the *R*-branch photon. Although population inversion between the excited state $B^2\Sigma_u^+$ and the ground state $X^2\Sigma_g^+$ of the nitrogen ions is created by the 800 nm pump pulses, there are still a portion of nitrogen ions remaining in the ground state. As a consequence, the rotational wave packet will be formed not only in the excited state $B^2\Sigma_u^+$ of the nitrogen ions, but also in the ground state $X^2\Sigma_g^+$. For facilitate the discussion, here we explicitly illustrate the two channels for generation of the nitrogen lasing in Fig. 6.

The first channel is a direct channel as illustrated in Fig. 6(a), in which a nitrogen molecule is first ionized into the excited state $B^2\Sigma_u^+$ $|J\rangle$ by the pump pulses, and then it will make a transition to the ground state $X^2\Sigma_g^+$ $|J-1\rangle$ by seeding with the probe pulses, leading to the emission of an *R*-branch photon. The transition probability of the first channel is dependent on the population in $B^2\Sigma_u^+$ $|J\rangle$, which shows a fast modulation at a frequency of $(4J+6)B_B c$ owing to the quantum beat between the coupling of $B^2\Sigma_u^+$ $|J\rangle$ and $B^2\Sigma_u^+$ $|J+2\rangle$ states as discussed above (see Figs. 3 and 4).

The second channel is a cascade process which is analogous to a resonant Raman process, as illustrated in Fig. 6(b). In this case, the nitrogen ion is initially in the ground state $X^2\Sigma_g^+$ $|J+1\rangle$, then it absorbs a *P*-branch photon and make a transition to the excited state $B^2\Sigma_u^+$ $|J\rangle$, finally it will return back to the ground state $X^2\Sigma_g^+$ $|J-1\rangle$ and emits a *R*-branch photon. Thus, the transition probability of the second channel relies on both the ground state $X^2\Sigma_g^+$ $|J+1\rangle$ and the excited state $B^2\Sigma_u^+$ $|J\rangle$. As discussed above, the population in the excited state $B^2\Sigma_u^+$ $|J\rangle$ will be modulated at a frequency of $(4J+6)B_B c$ due to the similar mechanism in a double-pulse alignment experiment [24]. Likewise, the population in ground state $X^2\Sigma_g^+$ $|J+1\rangle$ will be modulated at a frequency of

$$\Omega_2 = \frac{E_{J+3,X} - E_{J+1,X}}{h} = [4(J+1)+6]B_X c \qquad (3)$$

where $B_X$ = 1.92 cm$^{-1}$ [23] is the rotational constant of the electronic state $X^2\Sigma_g^+$. As a consequence, the laser emission originated from the second channel is modulated as the beat between the excited state $B^2\Sigma_u^+$ $|J\rangle$ and the ground state $X^2\Sigma_g^+$ $|J+1\rangle$. The beating frequency can be written as

$$\Omega_3 = |\Omega_1 - \Omega_2| = |(4J+6)B_B c - [4(J+1)+6]B_X c|, \tag{4}$$

To compare with the experimental observation, we plotted the theoretically calculated slow oscillation curve based on this beat frequency with the red dashed line in Fig. 5. The comparison with the measured data shows a perfect agreement.

## 5. Conclusion

In conclusion, we have directly observed the ultrafast dynamics of the rotational state distribution in a molecular wave packet using the nitrogen lasing emission. We reveal two channels in generation of the *R*-branch lines in the nitrogen laser. The first channel is a direct channel, in which a beat in the lasing signal has been observed as a result of the coupling between two neighboring $B^2\Sigma_u^+$ $|J\rangle$ and $B^2\Sigma_u^+$ $|J+2\rangle$ states. The second channel is a cascade process which is analogous to a resonant Raman process. In this case, the nitrogen ion is initially ionized into the ground state $X^2\Sigma_g^+$ $|J+1\rangle$, and then is pumped to the excited state $B^2\Sigma_u^+$ $|J\rangle$ by absorbing a P-branch photon, which can also generate the population inversion. Therefore, our finding sheds new light on the underlying mechanism of strong-field-ionization-induced air lasing. More generally speaking, the technique provides a useful tool for rotational molecular spectroscopy.

**ACKNOWLEDGMENTS**

This work was supported by National Basic Research Program of China (Grants No. 2011CB808100, No. 2014CB921300), National Natural Science Foundation of China (Grants No. 11127901, No. 11134010, No. 11204332, No. 11304330, and No. 11004209), Program of Shanghai Subject Chief Scientist (Grant No. 11XD1405500).

**Appendix: Electronic Spectroscopy of Diatomic Molecules**

Here we present the basic concept of electronic spectroscopy of diatomic molecules which is describe in [25]. According to the Born-Oppenheimer approximation, the total energy of a diatomic molecular is given by,

$$\tilde{E}_{total} = \tilde{v}_{el} + G(v) + F(J)$$

$$G(v) = \omega_e \left(v + \frac{1}{2}\right) - \omega_e \chi_e \left(v + \frac{1}{2}\right)^2 \quad \text{(A1)}$$

$$F(J) = BJ(J+1) - D[J(J+1)]^2$$

where $\tilde{v}_{el}$ is the energy at the minimum of the electronic potential energy curve. The selection rule for vibronic transition (vibrational transitions in electronic spectra) allows $\Delta v$ to take on any integral value. An electronic transition is made up of vibrational bands, and each band is in turn made of rotational transitions. The vibrational bands are labeled as $v' - v''$ (single prime denotes upper states and double primes lower states), and an electronic transition is often called a band system.

For singlet-singlet electronic transitions, rovibronic line intensities are determined by the populations of the levels, the Franck-Condon factors and the rotational line strength factor $S_{J''}$. In particular, the total power $P_{J'J''}$ (in $W/m^2$) emitted by an excited rovibronic state $|nv'J'\rangle$ is

$$P_{J'J''} = \frac{16\pi^3}{3\varepsilon_0 c^3} \frac{n_{J'}}{2J'+1} v^4 q_{v'v''} |R_e|^2 S_{J''}, \quad \text{(A2)}$$

in which $n_{J'}$ is the excited state population in molecules per $m^3$, $v$ is the transition frequency in Hz, $q_{v'v''}$ is the Franck-Condon factor, $R_e$ is the electronic transition dipole moment in coulomb meters and $S_{J''}$ is the rotational line strength factor.

$^1\Sigma - {}^1\Sigma$ transitions are referred to as parallel transitions, with the transition dipole moment lying along the z-axis.

For the $B^2\Sigma_u^+ \to X^2\Sigma_g^+$ transition of $N_2^+$, $\Delta\Lambda = 0$, $\Lambda'' = \Lambda' = 0$, transition has only P ($\Delta J = 1$) and R($\Delta J = -1$) branches. The rotational line strength factor is

$$S_J^R = \frac{(J''+1+\Lambda'')(J''+1-\Lambda'')}{J''+1} \tag{A3}$$

$$S_J^P = \frac{(J''+\Lambda'')(J''-\Lambda'')}{J''}.$$

**Figure captions:**

Fig. 1 (Color online) A typical spectrum of the 391 nm nitrogen lasing emission.

Fig. 2 (Color online) (a) The intensity of the total emission of *P*-branch band as a function of the time delay between the pump and the probe pulses. (b) The Fourier-transform of the curve in (a).

Fig. 3 (Color online) The signal intensities of the *R*-branch lines presented individually as a function of the time delay between the pump and probe pulses. Insets show the details in a window of time delay between 5 ps and 10 ps.

Fig. 4 (Color online) The Fourier-transforms of the corresponding curves in Fig. 3.

Fig. 5 (Color online) Experimental measured data (stars) and theoretical curves (dashed line) of the slow oscillation frequency as a function of rotational quantum number.

Fig. 6 (Color online) Energy level diagrams of the lasing emissions from (a) the direct channel, and (b) the resonant Raman-like cascade channel.

Fig. 1

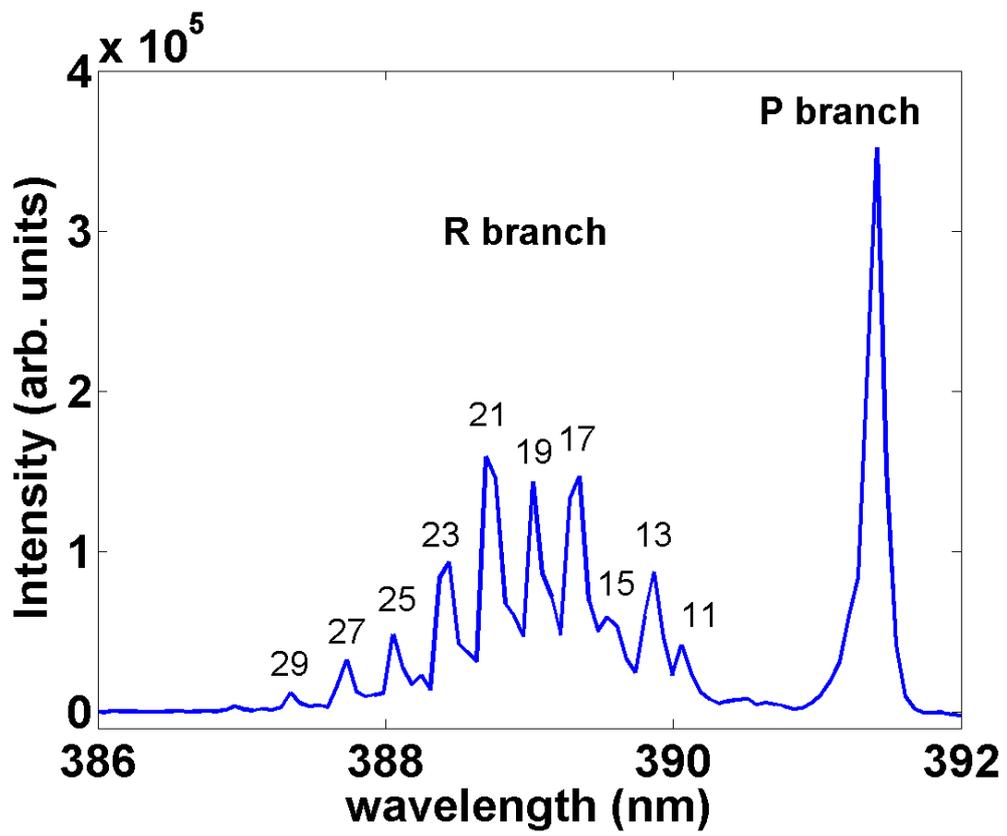

Fig. 2

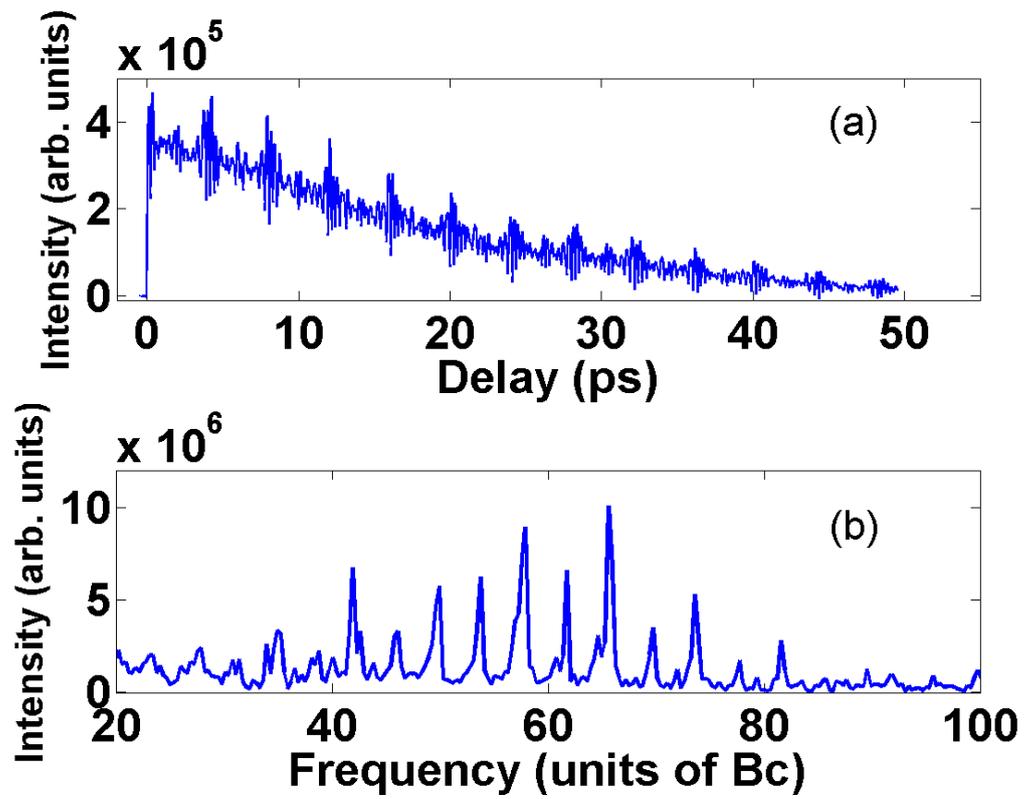

Fig. 3

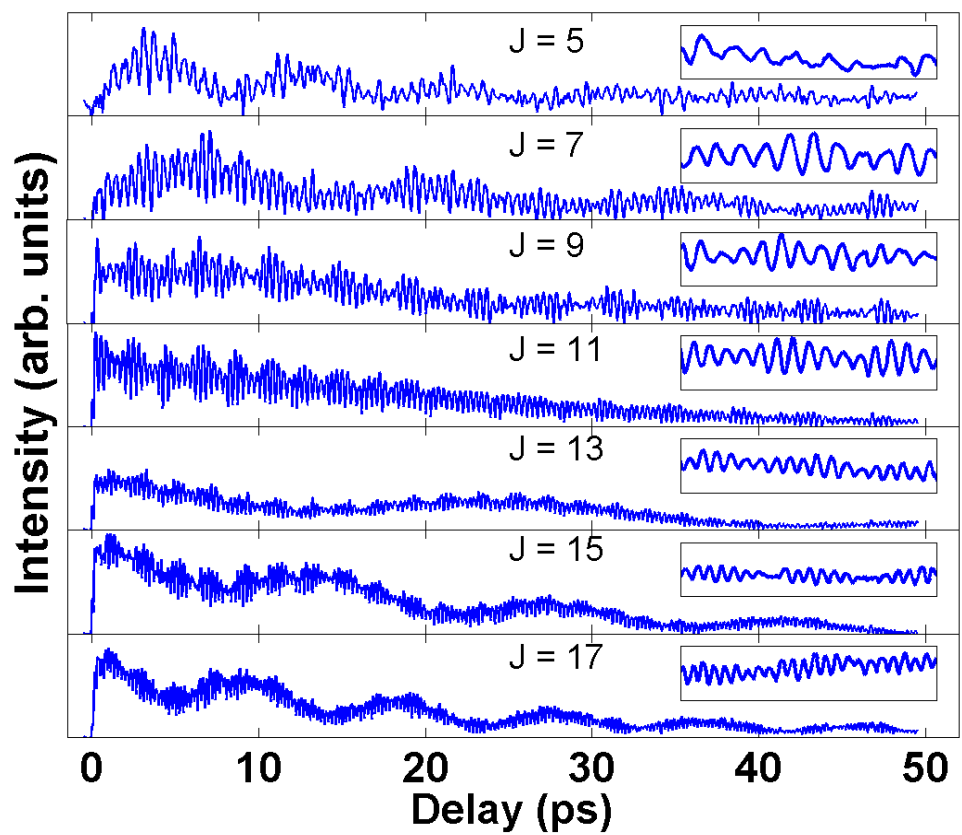

Fig. 4

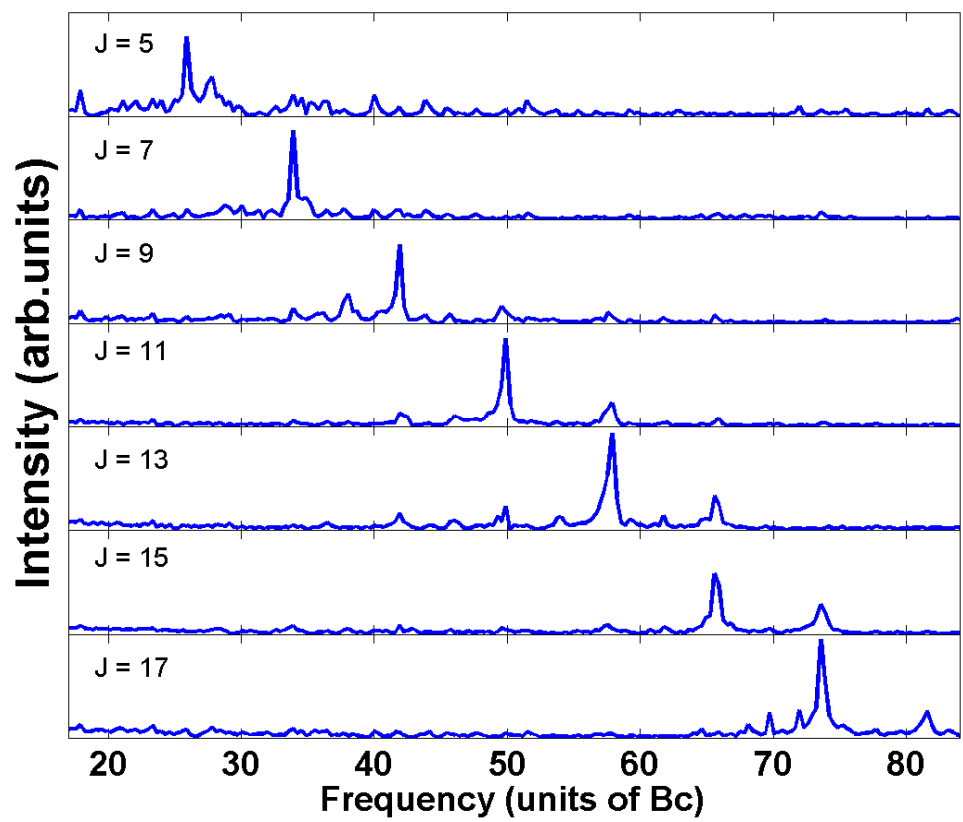

Fig. 5

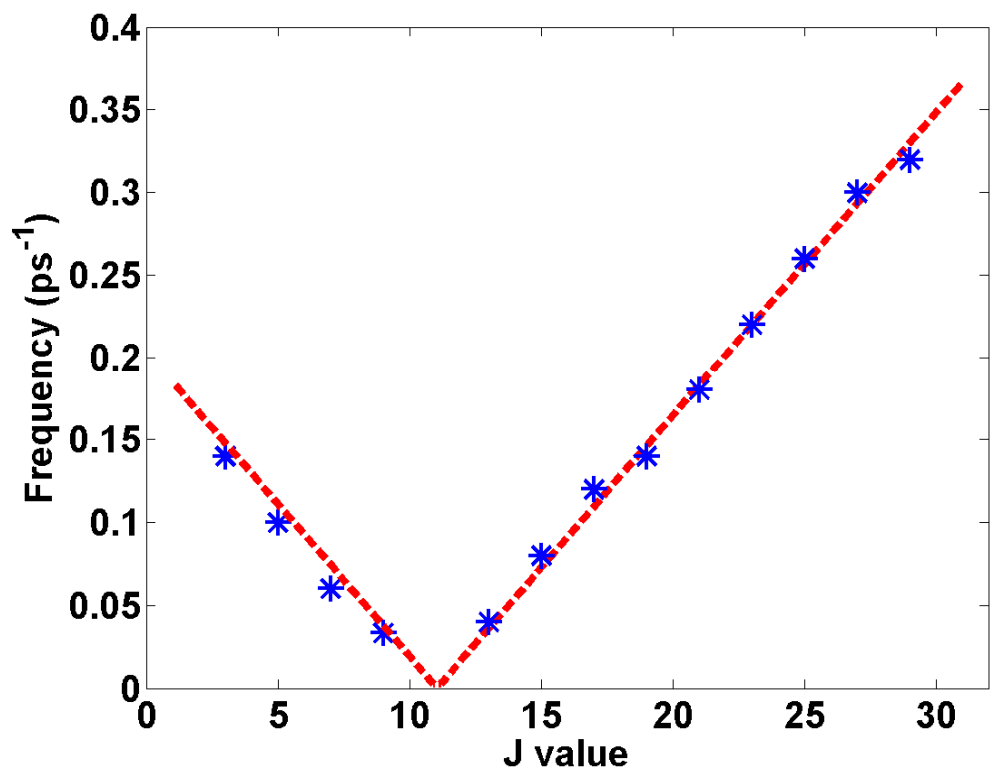

Fig. 6

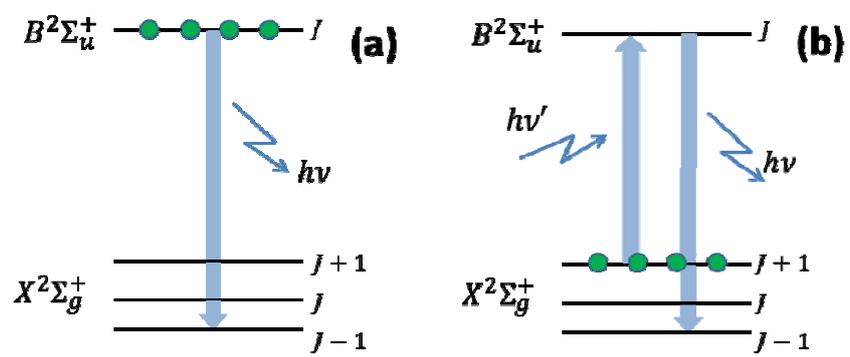